\documentclass[doublecol]{epl2}

\usepackage[utf8]{inputenx}
\usepackage{amsmath}
\usepackage{amsfonts}
\usepackage{amssymb}
\usepackage{lmodern}
\usepackage{cite}
\usepackage{epstopdf}
\usepackage{import}
\usepackage{bbold}
\usepackage{braket}
\usepackage{booktabs}
\usepackage{adjustbox}

\newcommand{\ii}{\mathrm{i}}
\renewcommand{\vec}[1]{\boldsymbol{#1}}

\newcommand{\num}{}
\renewcommand{\Re}{\mathrm{Re}\,}
\renewcommand{\Im}{\mathrm{Im}\,}

\title{Exceptional points in the elliptical three-disk scatterer using semiclassical periodic orbit quantization}
\shorttitle{Exceptional points in the elliptical three-disk scatterer}

\author{Niklas Liebermann \and Jörg Main \and Günter Wunner}
\shortauthor{Niklas Liebermann \etal}

\institute{
  1. Institut für Theoretische Physik, Universität Stuttgart - 70550 Stuttgart, Germany
}

\abstract{ The three-disk scatterer has served as a paradigm for semiclassical periodic orbit quantization of classical chaotic systems using Gutzwiller's trace formula. It represents an open quantum system, thus leading to spectra of complex eigenenergies. An interesting general feature of open quantum systems described by non-Hermitian operators is the possible existence of exceptional points where not only the complex eigenvalues but also their respective eigenvectors coincide. Using Gutzwiller's periodic orbit theory we show that exceptional points exist in a three-disk scatterer if the system's geometry is modified by extending the system from circular to elliptical disks. The extension is implemented in such a way that the system's characteristic $C_{3\mathrm{v}}$~symmetry is preserved. The two-dimensional parameter plane of the system is then spanned by the distance between and the excentricity of the elliptical disks. As typical signatures of exceptional points we observe the permutation of two resonances when an exceptional point is encircled in parameter space, and a non-Lorentzian resonance line shape in the weighted density of states.}

\begin{document}

\maketitle

\section{Introduction}
\begin{figure}[t]
  \centering
  \includegraphics{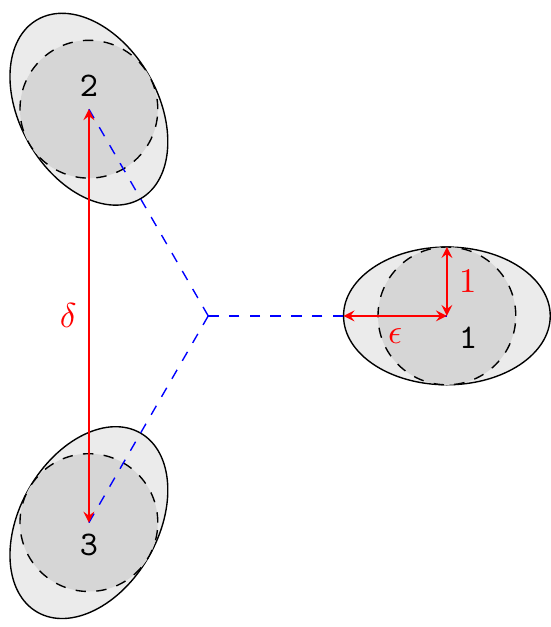}
  \caption{Potential landscape of the original three-disk scatterer with circles (dashed lines) and extension to ellipses (solid lines) to introduce the real parameters~$\delta$ (distance between the ellipses) and $\epsilon$ (semimajor axes of the ellipses).}
  \label{fig:sketch}
\end{figure}

Gutzwiller's trace formula which was published in 1971 allows for the semiclassical quantization of non-integrable classical systems~\cite{gutzwiller_phaseintegral_1967,gutzwiller_periodic_1971,gutzwiller_chaos_2013}. For the calculation of the semiclassical density of states, it uses classical periodic orbit data. The semiclassical approximation of the density of states~$\rho^\mathrm{scl}$ can then be divided into a mean part~$\bar{\rho}$ and a fluctuating part~$\rho_\mathrm{po}$ that is calculated by means of the classical periodic orbit data only. One of the most well-studied paradigms for the application of Gutzwiller's trace formula is the three-disk scattering system~\cite{eckhardt_fractal_1987,gaspard_scattering_1989}. It represents an open billiard quantum system with three hard disks (see dashed lines in fig.~\ref{fig:sketch}). The goal is to calculate the eigenenergies of the system using classical periodic orbit data, and subsequently evaluating Gutzwiller's trace formula. Since the trace formula consists of an infinite sum over all periodic orbits, it suffers from convergence problems. Several methods have been presented to resolve this issue. They include cycle expansion~\cite{cvitanovic_periodic-orbit_1989}, pseudo-orbit expansion~\cite{berry_rule_1990}, Padé resummation~\cite{main_semiclassical_1999}, and harmonic inversion analysis~\cite{main_use_1999}.

The resonances of open quantum systems are the eigenvalues of non-Hermitian operators and therefore complex numbers~\cite{moiseyev_non-hermitian_2011}. The real part of the eigenvalue represents the energy and the imaginary part is the decay rate of the respective state. An interesting feature of open quantum systems is the possible existence of exceptional points (EPs)~\cite{moiseyev_non-hermitian_2011,kato_perturbation_2013,heiss_phases_1999,heiss_physics_2012}. As opposed to simple degeneracy of resonances, both the eigenvalues and their respective eigenvectors coalesce. To observe EPs, there have to be at least two real-valued free parameters.

EPs have been experimentally shown in open billiard systems realized in microwave cavities~\cite{dembowski_experimental_2001,dietz_exceptional_2011,bittner_scattering_2014}. In addition, EPs have been theoretically shown to occur in atomic~\cite{magunov_strong_1999-1,magunov_laser-induced_2001-1,latinne_laser-induced_1995-1,cartarius_exceptional_2007-1} and molecular~\cite{lefebvre_resonance_2009} spectra, in scattering of particles at potential barriers~\cite{hernandez_non-hermitian_2006}, in atom waves~\cite{rapedius_nonlinear_2010,cartarius_discovery_2008,gutohrlein_bifurcations_2013,abt_supersymmetric_2015}, and in open Bose--Hubbard systems~\cite{graefe_non-hermitian_2008}. They appear in unstable lasers~\cite{berry_mode_2003}, resonators~\cite{klaiman_visualization_2008}, and optical waveguides~\cite{wiersig_asymmetric_2008,wiersig_enhancing_2014} as well. They have also been experimentally verified in metamaterials~\cite{lawrence_manifestation_2014}, a photonic crystal slab~\cite{zhen_spawning_2015}, electronic circuits~\cite{stehmann_observation_2004}, and in a chaotic exciton--polariton billiard system~\cite{gao_observation_2015}.

The question in what follows is whether exceptional points can be revealed by semiclassical quantization using Gutzwiller's trace formula. The three-disk scatterer will be investigated as a model system. Since the only modified parameter in preceding studies has been the distance between the disks, the system has to be extended by a second real-valued parameter. Therefore, the disks are given elliptical form where the second parameter is the length of the semimajor axis~$\epsilon$. The modification is performed in such a way that the system's characteristic $C_{3\mathrm{v}}$ symmetry is preserved. Harmonic inversion is used as an evaluation method to calculate both the resonance positions in the complex plane and their respective amplitudes. This allows for not only verifying the presence of exceptional points by encircling in parameter space and observing the permutation of two resonances~\cite{moiseyev_non-hermitian_2011} but also analyzing resonance shapes which can be non-Lorentzian~\cite{fuchs_harmonic_2014}.

\section{Theory}
In the first part of this section, the general ideas of Gutzwiller's periodic orbit theory and its applications to billiard systems are discussed. In the second part, the theoretical foundations of exceptional points in open quantum systems will be outlined briefly.

\subsection{Periodic orbit theory}
\begin{figure}
  \centering
  \includegraphics{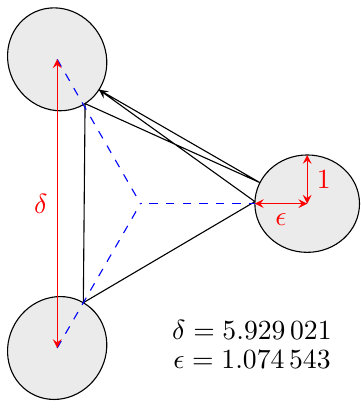}
  \caption{Periodic orbit for the symbolic code \texttt{1231212} and for the parameters~$\delta$ and $\epsilon$ given. The periodic orbits are found by minimizing the length of the respective orbit.}
  \label{fig:orbits_example}
\end{figure}

The density of states can be split into a mean density~$\bar{\rho}$ and an oscillating part~$\rho_\mathrm{po}$ in a semiclassical approximation since there are additional contributions close to periodic orbits in the semiclassical limit $\hbar \to 0$. The mean density of states as a function of the energy~$E$ is given by the Thomas--Fermi term. The oscillating part is given by Gutzwiller's trace formula
\begin{equation}
  \rho_\mathrm{po}(E) = \frac{1}{\pi \hbar} \sum_\mathrm{po} \frac{T_\mathrm{ppo}}{\lvert \det M_\mathrm{po} - \mathbb{1}\rvert} \cos\bigg(\frac{S_\mathrm{po}}{\hbar} - \frac{\pi}{2} \sigma_\mathrm{po}\bigg)\,,
  \label{eq:gutzwiller}
\end{equation}
where the summation is over all classical periodic orbits, including multiple repetitions of the same primitive periodic orbit. Here, $S_\mathrm{po}$ is the action along the respective periodic orbit, $M_\mathrm{po}$ is the monodromy matrix, $\sigma_\mathrm{po}$ denotes the Maslov index, and $T_\mathrm{ppo}$ is the period of a primitive periodic orbit.

The first step to calculating the semiclassical density of states of the elliptical three-disk scatterer is the evaluation of the classical periodic orbit dynamics. Every periodic orbit can be described by means of a symbolic code which gives the sequence in which the disks \texttt{1}, \texttt{2}, and \texttt{3} are traversed~\cite{gaspard_exact_1989}. Since the system is subject to a $C_{3\mathrm{v}}$~symmetry, a symmetry-reduced symbolic code can be introduced~\cite{cvitanovic_periodic-orbit_1989}. The symmetry reduction based on the system's $C_{3\mathrm{v}}$ symmetry can still be used when the extension to ellipses is done with the semimajor axes all pointing towards the centre of symmetry (see fig.~\ref{fig:sketch}). The two parameters to find exceptional points in the system are the distance~$\delta$ of the elliptical disks and the length of the semimajor axes~$\epsilon$. The length of the semiminor axes is kept constant at~$1$.

The periodic orbits of the elliptical three-disk scatterer can be found by minimizing the geometric length of a trajectory associated with a particular symbolic code. Figure~\ref{fig:orbits_example} shows an example for a minimized trajectory for a specific set of parameters~$\delta$ and~$\epsilon$ and one specific symbolic code. The required data to evaluate Gutzwiller's trace formula~\eqref{eq:gutzwiller} in a billiard system are the orbit length, the monodromy matrix, and the Maslov index. Since the potential energy is zero between the disks, only the time of flight between the disks and the collisions contribute to the monodromy matrix of a trajectory~\cite{bogomolny_smoothed_1988}.
The Maslov index increases by~$2$ for every reflection at a hard wall. It is therefore given by $\sigma_\mathrm{po} = 2 \ell_\mathrm{po}$ with~$\ell_\mathrm{po}$ being the length of the symbolic code of the periodic orbit~\cite{eckhardt_quantum_1995}.

Gutzwiller's trace formula~\eqref{eq:gutzwiller} represents a sum over infinitely many primitive periodic orbits and an infinite amount of repetitions of all of these. Directly evaluating the sum for periodic orbits up to a certain geometric or symbolic length leads to convergence problems. Several methods have been developed to improve the convergence of Gutzwiller's trace formula. The most favourable method for the problem at hand is harmonic inversion since it allows for the calculation of expectation values of quantum mechanical operators. To observe the special features exhibited by the resonance amplitudes close to an exceptional point, we use diagonal matrix elements of quantum mechanical observables for the weighted density of states which can be calculated using an extension of Gutzwiller's trace formula.

For this purpose, we start from the quantum mechanical response function
\begin{equation}
  g^\mathrm{qm} = \lim_{\epsilon \to 0} \sum_n \frac{\braket{n|\hat{D}|n}}{E-E_n+\mathrm{i}\epsilon}
\end{equation}
with the requirement, that
\begin{equation}
  \rho_{\hat{D}} = -\frac{1}{\pi} \Im g^\mathrm{qm}
  \label{eq:weighted_density}
\end{equation}
is the weighted density of states with the residues being the diagonal matrix elements of an observable~$\hat{D}$~\cite{eckhardt_semiclassical_1992}. For real eigenenergies, $\rho_{\hat{D}}$ is a sum over delta functions with each summand contributing with weight~$\braket{n|\hat{D}|n}$. For complex eigenenergies, $\rho_{\hat{D}}$ consists of Lorentzian functions where the complex matrix elements~$\braket{n|\hat{D}|n}$ determine the Lorentzian line shape.

In Gutzwiller's semiclassical periodic orbit approximation, the semiclassical response function~$g^\mathrm{scl}$ is the sum of the two parts~$\bar{g}^\mathrm{scl}$ -- the mean part of the response function associated with the Thomas--Fermi term -- and~$g^\mathrm{scl}_\mathrm{po}$ -- the oscillating periodic orbit contribution associated with eq.~\eqref{eq:gutzwiller}. For billiard systems, the semiclassical approximation for the matrix elements of a quantum mechanical operator leads to an oscillating part of the semiclassical response function
\begin{equation}
  g^\mathrm{scl}_\mathrm{po}(k) = \sum_\mathrm{po} D_\mathrm{po} \underbrace{\frac{T_\mathrm{ppo}(-1)^{\ell_\mathrm{po}}}{\bigl\lvert \det (M_\mathrm{po} - \mathbb{1})\bigr\rvert}}_{\equiv \mathcal{A}_\mathrm{po}} \exp(\ii k L_\mathrm{po})
  \label{eq:scl_billiard_response}
\end{equation}
with the amplitudes~$\mathcal{A}_\mathrm{po}$ and the wave number $k = \sqrt{2mE}/\hbar$. $L_\mathrm{po}$ is the length of the periodic orbit, which is equivalent to the action~$k L_\mathrm{po}$ along the trajectory in a billiard system. The additional factor~$D_\mathrm{po}$ is given by
\begin{equation}
  D_\mathrm{po} = \frac{1}{T_\mathrm{po}} \int_0^{T_\mathrm{po}} D\bigl(\vec{q}(t), \vec{p}(t)\bigr)\,\mathrm{d}t
  \label{eq:mean_value}
\end{equation}
which is the average of the classical observable~$D$ along the trajectory of a periodic orbit. The phase space variable~$D$ is the Wigner transform of the quantum mechanical operator~$\hat{D}$~\cite{eckhardt_semiclassical_1992}.

The Fourier transform of the semiclassical response function~\eqref{eq:scl_billiard_response} yields the semiclassical signal~$C^\mathrm{scl}(L)$. The idea of harmonic inversion entails adapting the exact quantum mechanical expression of the signal to the semiclassical signal that can be calculated from classical periodic orbit data only. Convergence can be further improved by band-limiting the signal to a certain frequency window~$[k_0 - \Delta k, k_0 + \Delta k]$ to be analyzed~\cite{wall_extraction_1995,mandelshtam_spectral_1997,mandelshtam_harmonic_1997,main_decimation_2000}. The calculation of the complex resonance positions~$k_n$ and their respective complex amplitudes~$d_n$, that correspond to the diagonal matrix elements~$\braket{n|\hat{D}|n}$, then boils down to calculating the semiclassical band-limited signal for periodic orbits up to a certain maximum periodic orbit length~$L_\mathrm{max}$
\begin{equation}
  C^\mathrm{scl}_\mathrm{bl}(L) = \sum_\mathrm{po} D_\mathrm{po} \mathcal{A}_\mathrm{po} \frac{\sin\big[ (L - L_\mathrm{po}) \Delta k \big]}{\pi(L - L_\mathrm{po})} \exp(\ii L_\mathrm{po} k_0)
  \label{eq:scl_signal}
\end{equation}
on an equidistant grid with points~$L_i$. The semiclassical signal is subsequently adapted to the exact quantum mechanical band-limited signal
\begin{equation}
  C^\mathrm{qm}_\mathrm{bl}(L) = -\ii \sum_{n=1}^N d_n \exp\big[{-\ii} (k_n - k_0) L \big]\,,
  \label{eq:qm_signal}
\end{equation}
which consists of a finite number~$N$ of frequencies in the same frequency window, and evaluated on the same grid~$L_i$. To solve the resulting nonlinear set of equations methods like the linear predictor, Padé approximation, and signal diagonalization can be used~\cite{main_decimation_2000}. For all following calculations the Padé approximant will serve as a method of solution.

\subsection{Exceptional points}
Open quantum systems can most efficiently be described by non-Hermitian Hamiltonians. This description leads in particular to complex resonances. The decay rate of unbound states is then included in the imaginary part, thus avoiding difficult-to-treat time-dependent expressions. Systems with complex resonances exhibit special features, especially in the vicinity of an exceptional point (EP). Contrary to Hermitian systems where in case of degeneracy only the eigenvalues coincide, in a non-Hermitian system the respective eigenstates also coalesce at an EP.

The most simple model of a system with an EP is the two-dimensional matrix
\begin{equation}
  H(\lambda) = \begin{pmatrix}
    1 & \lambda\\
    \lambda & -1
  \end{pmatrix}
\end{equation}
that depends on a complex parameter~$\lambda$~\cite{kato_perturbation_2013}. The matrix is non-Hermitian obviously for non-real~$\lambda$. The eigenvalues are given by $\epsilon_1 = \sqrt{1+\lambda^2}$ and $\epsilon_2 = -\sqrt{1+\lambda^2}$. They become degenerate for $\lambda_0 = {\pm \ii}$. The same holds true for the respective eigenvectors. Therefore, $\lambda_0$ is an EP. One of the most striking features of an EP can be observed when the EP is encircled in parameter space, \emph{e.\,g.} $\lambda(\phi) = \ii + r \exp(\ii\phi)$ for $\phi = 0 \dots 2\pi$. The eigenvalues in a power-series expansion are then given by $\epsilon_1 = \sqrt{2r} \exp\bigl[\ii (\pi/4 + \phi/2)\bigr]$ and $\epsilon_2 = \sqrt{2r} \exp\bigl[\ii (5\pi/4 + \phi/2)\bigr]$. One can see that for small encircling radii~$r$ the eigenvalues interchange their positions when the EP is encircled in a closed loop in parameter space. The EP has to be encircled twice to make the resonances return to their original positions. However, the respective eigenvectors pick up an additional phase of~$\pi$. This is a characteristic feature of an EP and will be used as a distinct identifier for an EP in the elliptical three-disk scatterer.

In addition, EPs show up as poles of higher orders in the response function~$g$.
Non-degenerate resonances at positions~$k_n$ manifest themselves in the quantum mechanical response function~$g^\mathrm{qm}$ as a sum over first-order poles as
\begin{equation}
  g^\mathrm{qm} = \sum_n \frac{\braket{n|\hat{D}|n}}{k-k_n}\,.
  \label{eq:g_nondegenerate}
\end{equation}
This ansatz was used to derive the quantum mechanical signal in eq.~\eqref{eq:qm_signal}. However, if there are degenerate resonances in the spectrum, the expectation values for those specific resonances diverge. The ansatz from eq.~\eqref{eq:g_nondegenerate} is not sufficient anymore. Instead, the response function has to be generalized to
\begin{equation}
  g^\mathrm{qm} = \sum_n \sum_{\alpha = 1}^{r_n} \frac{d_{n, \alpha}}{(k - k_n)^\alpha}\,,
  \label{eq:g_degenerate}
\end{equation}
where $r_n$ is the order of degeneracy of the $n$-th resonance. The amplitudes $d_{n, \alpha}$ do not show divergence at the EP anymore. In this case, the quantum mechanical signal from eq.~\eqref{eq:qm_signal} has to be generalized to
\begin{equation}
  C^\mathrm{qm}_\mathrm{bl}(L) = -\ii \sum_{n=1}^N \sum_{\alpha = 1}^{r_n} d_{n, \alpha} \frac{(-\ii L)^{\alpha - 1}}{(\alpha - 1)!} \exp\bigl[-\ii (k_n - k_0) L\bigr]
\end{equation}
and the Padé approximant has to be adapted~\cite{fuchs_harmonic_2014}. The higher-order contributions of the difference~$(k-k_n)$ lead to a non-Lorentzian shape of the resonances in the weighted density of states. This is another indicator for an EP that can be observed.

\section{Results}
\begin{figure}[t]
  \centering
  \includegraphics{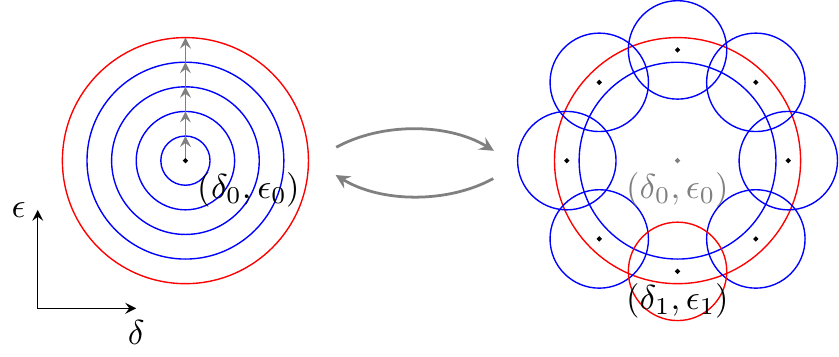}
  \caption{Procedure to find an exceptional point by observing state permutation. The procedure starts from an arbitrary point~$(\delta_0, \epsilon_0)$. The radius of the encircling in parameter space is gradually increased (left figure). As soon as a permutation of states is observed in resonance space, encirclings with smaller radii are performed along the outer ring (right figure). In at least one of the small circles, a permutation of states will be observed again. By iterating this procedure, the EP can in principle be localized up to an arbitrary precision.}
  \label{fig:ep_radius_increase}
\end{figure}

\begin{figure}[t]
  \centering
  \includegraphics{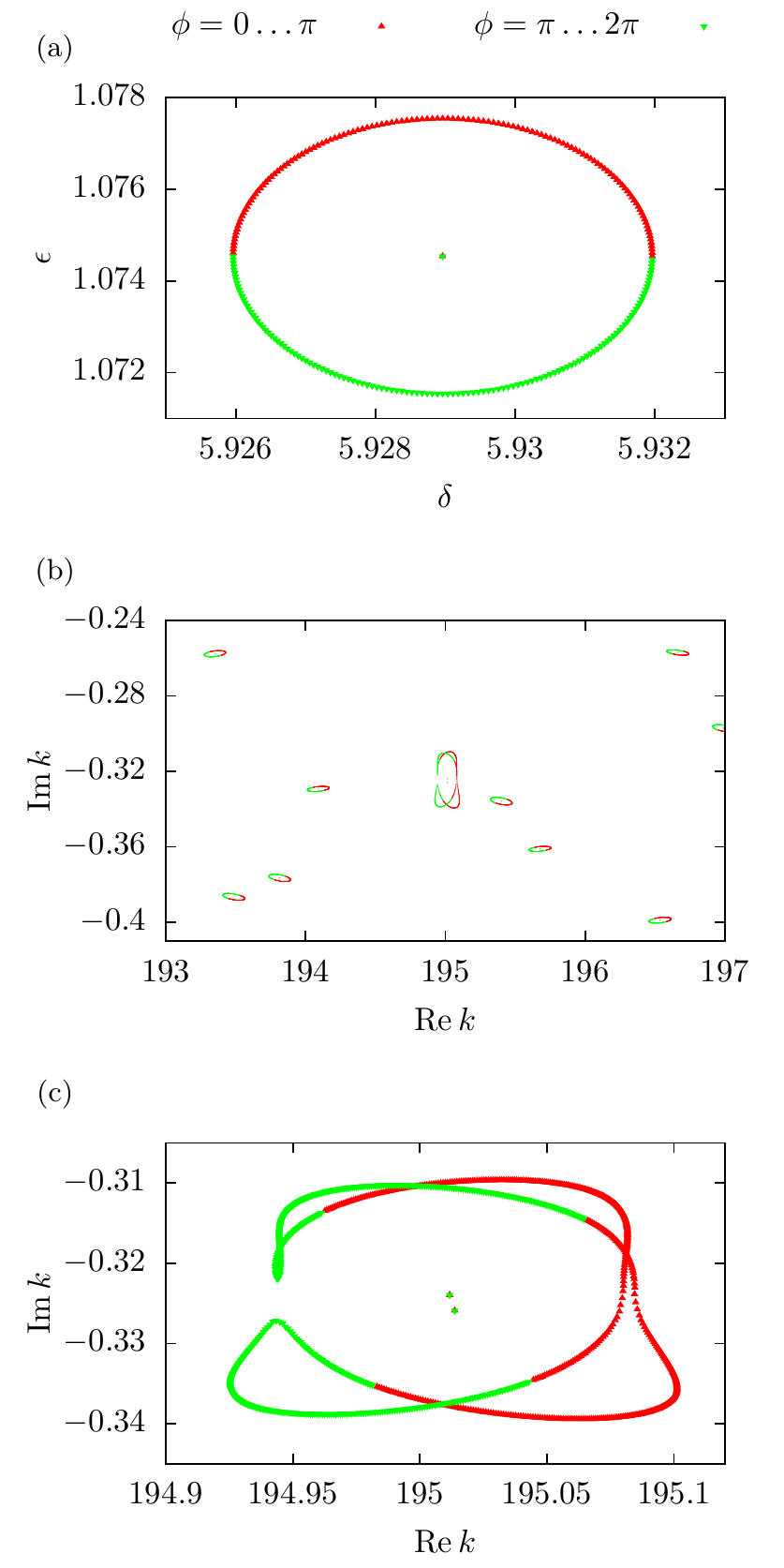}
  \caption{Encircling of an exceptional point both in parameter space~$(\delta, \epsilon)$ with centre~$(\delta_0, \epsilon_0)$ (a) and in energy space~$(\Re k, \Im k)$ for selected resonances (b, c). To better illustrate the permutation of states, the first half of the encircling ($0 \dots \pi$) is coloured in red, whereas the second half is ($\pi \dots 2\pi$) is coloured in green. Non-degenerate resonances simply return to their original position after one full circle. Only at a position where two resonances are close in the complex plane, the spectrum shows two resonances switching positions when encircling the EP.}
  \label{fig:ep_orbit}
\end{figure}

The spectrum of the three-disk scatterer with circles can be depicted in the complex plane. It consists of isolated resonances. The spectrum can be changed continuously by introducing the real-valued parameters~$\delta$ and~$\epsilon$ (see fig.~\ref{fig:sketch}). In this way, the resonances are moved in the complex plane to find a set of parameters $(\delta_0, \epsilon_0)$ where two of them coalesce at an exceptional point (EP) (see fig.~\ref{fig:ep_radius_increase}).

\subsection{Calculation of resonances}
To calculate the resonances of the extended three-disk scatterer, the signal in eq.~\eqref{eq:scl_signal} has to be evaluated up to a reasonable amount of periodic orbits. As already mentioned, the periodic orbit data necessary for the calculation are the geometric length of the periodic orbit (which corresponds to the action along the trajectory), the determinant of the monodromy matrix, and the Maslov index. The mondromy matrix of the minimized orbit is calculated by matrix multiplication of the contributions of each trajectory segment and reflection. The Maslov index~$\sigma_\mathrm{po}$ is twice the sequence length~$\ell_\mathrm{po}$. When inserted in the equation for the semiclassical band-limited signal~$C^\mathrm{scl}_\mathrm{bl}$, it leads to a prefactor of~$1$ for hyperbolic symbolic dynamics (even~$\ell_\mathrm{po}$) or~$-1$ for inverse hyperbolic symbolic dynamics (odd~$\ell_\mathrm{po}$).

Since the number of periodic orbits for every sequence length added grows exponentially and the precision of the calculated resonances only improves slightly above a certain sequence length, the maximum sequence length is chosen to be $\ell_\mathrm{max} = 15$.

With those parameters known for every periodic orbit up to~$\ell_\mathrm{max}$, the signal for band-limited harmonic inversion from eq.~\eqref{eq:scl_signal} can be calculated. Band-limited harmonic inversion allows for the exact calculation of resonances even for high $k$~numbers. In the search for exceptional points, a frequency window is chosen in a high $k$~range from~$k_\mathrm{min} = 175$ to~$k_\mathrm{max} = 205$.

\subsection{Search for exceptional points}
In order to find EPs in those regions, a point in the parameter plane spanned by~$\delta$ and~$\epsilon$ is encircled with semiaxes~$r_\delta$ along the $\delta$ axis and~$r_\epsilon$ along the $\epsilon$ axis. An encircling with $r_\delta = r_\epsilon = 0.003$ in parameter space is shown in fig.~\ref{fig:ep_orbit}a and is used for the following illustrations in the resonance spectra.

The corresponding resonances for every parameter pair along the circle are tracked in a predefined window in the resonance plane. If all resonances return to their original positions after a full circle from~$0$ to~$2\pi$, there is no indication of an EP. However, if an EP is contained in the circle, a permutation of a pair of resonances can possibly be observed. It has to be noted that for better illustration the circle is divided into two parts~$0 \dots \pi$ (red dots) and $\pi \dots 2\pi$ (green dots). If there is no permutation of states, there are only two colour changes in the circle in resonance space. However, if there is a permutation of states, the colours change four times during the course of one circle.

The iterative procedure to find an EP is sketched in fig.~\ref{fig:ep_radius_increase}. In a first step, the arbitrarily chosen initial point (in our example $\delta_0 = 6$ and $\epsilon_0 = 1$) is encircled with radii~$r_\delta$ and~$r_\epsilon$ that are increased in small steps. The respective spectra are calculated. If no EP is contained in the circle, all the resonances in the spectrum show only two colour changes, hence return to their original position after one circle. During the process of increasing~$r_\delta$ and~$r_\epsilon$, a fourfold colour change in the resonance spectrum may occur in a certain step~$i_0$. This indicates that an EP is located in the ring between the circles in step~$i_0 - 1$ and~$i_0$. The new centre points of the following circles, which can now be performed with considerably smaller radii, are placed along that ring. The amount of points along the ring has to be chosen in a way that such circles overlap. In at least one of the circles, the EP is located and a permutation of the two resonances will be observed again. This procedure can be iterated and allows for the determination of the EP position with arbitrary precision, in principle.

However, the precision of the calculations is limited by the amount of periodic orbits used for the semiclassical quantization. Since the number of orbits rises exponentially with symbol length, at some point the precision of the calculation only increases slightly. For a maximum symbol length of $\ell_\mathrm{max} = 15$ the calculation allows for a determination of the EP up to a precision of five digits. An illustration of the vicinity of the EP at ($\delta_\mathrm{EP} = \num{5.928\,96}, \epsilon_\mathrm{EP} = \num{1.074\,54}$) is given in fig.~\ref{fig:ep_orbit}b. Non-degenerate resonances do not show deformation when encircled and simply return to their original position after one full circle in parameter space. However, at a position where two resonances come close to each other, deformation and ultimately permutation of resonances is observed. A close-up of two resonances permuting in resonance space is clearly depicted in the enlargement in fig.~\ref{fig:ep_orbit}c.

\subsection{Analysis of resonance amplitudes}
\begin{figure}[t]
  \includegraphics{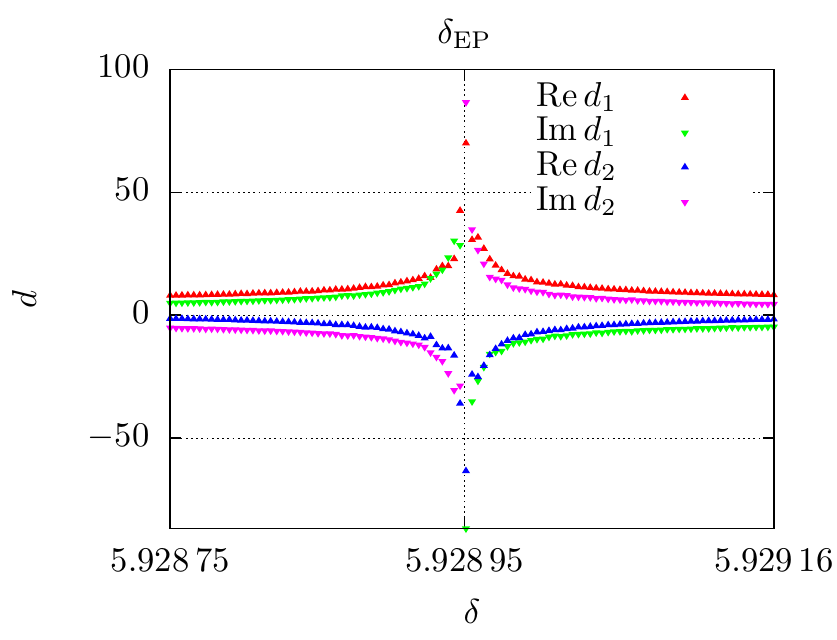}
  \caption{Amplitudes of the two EP resonances along a linear pathway along the $\delta$ axis in parameter space using ansatz~\eqref{eq:g_nondegenerate}. $\epsilon$ is held constant at $\epsilon = \epsilon_\mathrm{EP} = \num{1.074\,54}$. When approaching the EP at $\delta_\mathrm{EP} = \num{5.928\,96}$ the amplitudes rise exponentially with a divergence exactly at~$\delta_\mathrm{EP}$.}
  \label{fig:ep_linear_zdod1}
\end{figure}

\begin{figure}[t]
  \centering
  \includegraphics{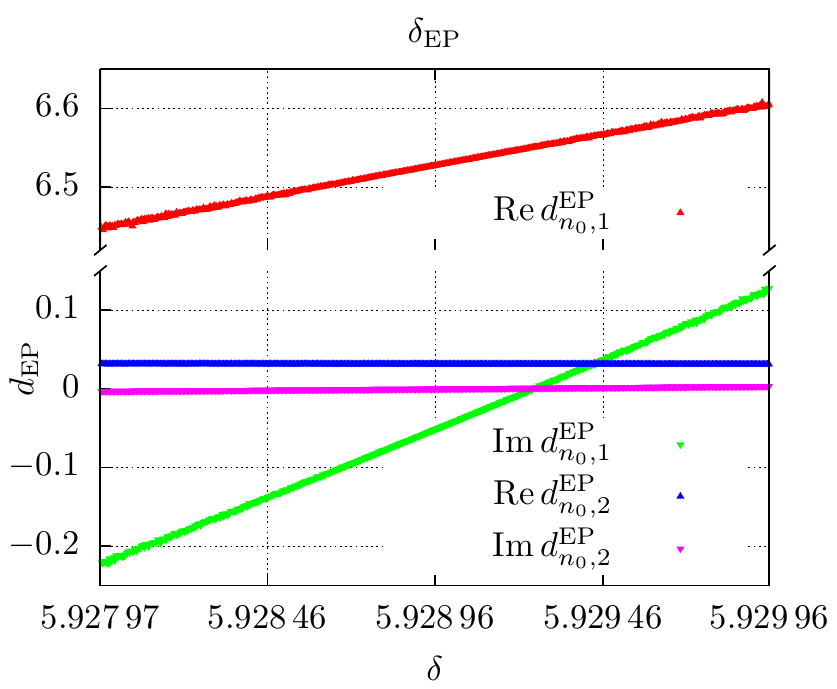}
  \caption{Semiclassically calculated diagonal matrix elements~$\braket{n_0, \alpha | \hat{\vec{r}}^2 | n_0, \alpha}$ along a linear pathway along the $\delta$ axis in parameter space ($\epsilon = \epsilon_\mathrm{EP}$) using the extended ansatz~\eqref{eq:g_degenerate}. The new amplitudes~$\hat{d}$ do not diverge at the EP. There is a non-zero contribution of the second-order term.}
  \label{fig:ep_linear}
\end{figure}

\begin{figure}[t]
  \centering
  \includegraphics{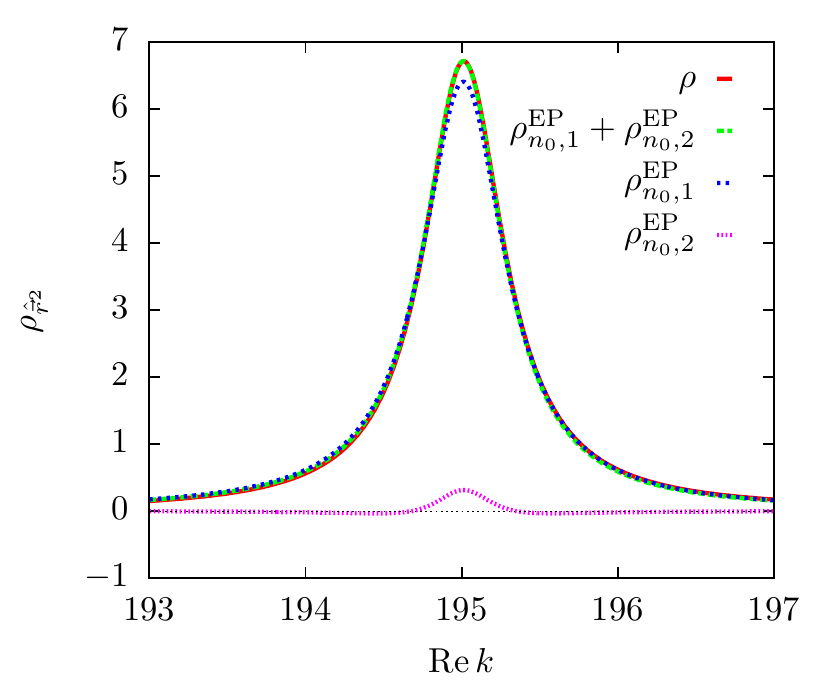}
  \caption{Line shape at the degenerate resonance in the weighted density of states for the operator~$\hat{\vec{r}}^2$. The combined line shape agrees for both ansatzes~\eqref{eq:g_nondegenerate} and~\eqref{eq:g_degenerate}. When analyzing the contributions from ansatz~\eqref{eq:g_degenerate} seperately, a non-zero contribution of the second-order term is noticeable, which implies that the combined line shape is non-Lorentzian.}
  \label{fig:linear_nonlorentz}
\end{figure}

In the density of states, the residues determine the degree of degeneracy of the respective eigenenergy. So when using Gutzwiller's trace formula without the extension for the calculation of diagonal matrix elements, the amplitudes are equal to~$1$ for every non-degenerate resonance as expected. Yet, when looking at the density of states weighted with diagonal matrix elements~$\braket{n|\hat{D}|n}$ as defined in eq.~\eqref{eq:weighted_density}, the amplitudes show a characteristic behaviour close to the EP. As an example, we will look at the classical observable~$D = \vec{r}^2$ with its quantum mechanically corresponding operator~$\hat{D} = \hat{\vec{r}}^2$. $\vec{r}^2$~then is the squared distance on the classical trajectory of a periodic orbit from the centre of symmetry (see fig.~\ref{fig:sketch}). The value of $\hat{\vec{r}}^2$ is invariant when the symmetry-reduced symbolic code is applied.

According to eq.~\eqref{eq:scl_signal}, the contribution to the semiclassical signal of each periodic orbit has to be multiplied by the mean value of the classical observable along a periodic orbit. The mean value~$r^2_\mathrm{po}$ is calculated by means of eq.~\eqref{eq:mean_value} as the classical expectation value along a periodic orbit. For the elliptical three-disk system it can be readily split into a sum over all trajectory segments.

Figure~\ref{fig:ep_linear_zdod1} depicts the behaviour of the amplitudes~$d_n = \braket{n | \hat{\vec{r}}^2 | n}$ of the two permuting resonances labelled~$n=1$ and~$n=2$ in the weighted density of states
\begin{equation}
  \rho_{\hat{\vec{r}}^2}  = -\frac{1}{\pi} \Im \sum_n \frac{\braket{n | \hat{\vec{r}}^2 | n}}{k-k_n}
\end{equation}
close to the EP. The EP $(\delta_\mathrm{EP}, \epsilon_\mathrm{EP})$ is traversed along the $\delta$~axis in parameter space while~$\epsilon$ is held constant at~$\epsilon_\mathrm{EP}$. The amplitudes show an exponential rise with a divergence directly at $\delta = \delta_\mathrm{EP}$. The divergence occurs because the ansatz for the quantum mechanical response function~\eqref{eq:g_nondegenerate} with only first-order poles is not valid close to an EP.

When using the modified ansatz~\eqref{eq:g_degenerate}, however, the behaviour of the residues~$d_{n,\alpha}$ in
\begin{equation}
  \rho^\mathrm{EP}_{\hat{\vec{r}}^2} = \sum_n \sum_\alpha \rho^\mathrm{EP}_{n, \alpha} = -\frac{1}{\pi} \Im \sum_n \sum_\alpha \frac{d^\mathrm{EP}_{n, \alpha}}{(k-k_n)^\alpha}
\end{equation}
is shown in fig.~\ref{fig:ep_linear} where the relevant resonance is labelled by~$n = n_0$. In the numerical calculations it is sufficient to consider two resonances as degenerate when they are close compared to the distance to all other resonances. The amplitudes now depend only weakly on the distance of the resonance and there is no divergence at the EP. Instead, there is a non-zero contribution of the second-order pole in the weighted density of states. The density of states~$\rho_{\hat{\vec{r}}^2}$ close to the EP position at $\Re k = \num{195.013}$ is depicted in fig.~\ref{fig:linear_nonlorentz}. The combined line shape calculated from the corrected ansatz~\eqref{eq:g_degenerate} (green line) agrees with the the one calculated from the first-order ansatz~\eqref{eq:g_nondegenerate}. Yet, when looking at the first-order (blue line) and second-order (pink line) term from ansatz~\eqref{eq:g_degenerate} individually, it is visible that there is a contribution of the second-order term of about~$10\,\%$. This clearly indicates that the combined line shape is non-Lorentzian.

\begin{table}[!b]
  \caption{Synopsis of the properties of the exceptional point.}
  \label{tab:conclusion}
  \begin{center}
    \begin{tabular}{lc}
    \toprule
      parameters & $\delta = \num{5.928\,96}, \epsilon = \num{1.074\,54}$\\
      \addlinespace[0.5ex]
      resonance position & $k = \num{195.013\,0} - \ii\,\num{0.324\,544}$\\
      \addlinespace[0.5ex]
      $\hat{\vec{r}}^2$ matrix elements & \begin{tabular}{@{}c@{}}$d_{n_0, 1} = \num{6.528\,114} - \ii\,\num{0.049\,424}$ \\ $d_{n_0, 2} = \num{0.031\,336} + \ii\,\num{0.000\,251}$
      \end{tabular}\\
    \bottomrule
    \end{tabular}
  \end{center}
\end{table}

\section{Conclusion}
We have extended the three-disk scatterer, which has been widely used as a paradigm for semiclassical periodic orbit quantization of an open quantum system, to the problem of finding exceptional points in the complex spectrum of eigenvalues. The extended system includes two real-valued parameters, the distance and the ellipticity of the disks. Using the characteristic permutation of states to identify EPs, they are searched for in the spectrum of the extended three-disk scatterer in an interative procedure. An EP is found close to the original configuration $(\delta_0 = 6, \epsilon_0 = 1)$. This leads to the expectation that there is an infinite amount of exceptional points in the elliptical three-disk scatterer. The properties of the EP are summarized in table~\ref{tab:conclusion}. For the analysis of diagonal matrix elements close to the EP, an operator~$\hat{\vec{r}}^2$ is defined. The analysis of the density of states weighted with the diagonal matrix elements reveals that -- when fitted to an ansatz not valid for degenerate resonances -- the amplitudes diverge at the EP as expected. When using an ansatz with higher-order terms the amplitudes show smooth, linear behaviour when scanned along the EP. There is a non-zero contribution of the second-order term in the weighted density of states close to the EP, leading to a non-Lorentzian line shape. Similar to the experiments previously performed in open billiard systems~\cite{dembowski_experimental_2001,dietz_exceptional_2011,bittner_scattering_2014}, microwave experiments could be used to experimentally verify the calculations by means of Gutzwiller's trace formula and confirm the occurence of exceptional points in the elliptical three-disk scatterer.


\raggedright

\end{document}